\documentclass[aps,prc,twocolumn,superscriptaddress,amsfonts,amsmath,amssymb,showpacs,floatfix]{revtex4}
\usepackage{dcolumn,multirow,epsfig}

\begin{document}

\title{Nuclear Structure Relevant to Neutrinoless Double $\beta$ Decay:\\
  the Valence Protons in $^{76}$Ge and $^{76}$Se.}

\author{B.P.~Kay}
\affiliation{Physics Division, Argonne National Laboratory, Argonne, IL 60439, USA}
\author{J.P.~Schiffer}
\affiliation{Physics Division, Argonne National Laboratory, Argonne, IL 60439, USA}
\author{S.J.~Freeman} \affiliation{University of Manchester, Manchester M13 9PL, United Kingdom}
\author{T.~Adachi}
\affiliation{Research Center for Nuclear Physics, Osaka University, Ibaraki, Osaka 567-0047, Japan}
\author{J.A.~Clark}
\affiliation{Yale University, New Haven, CT 06520, USA}
\author{C.M.~Deibel}
\affiliation{Yale University, New Haven, CT 06520, USA}
\author{H.~Fujita}
\affiliation{Research Center for Nuclear Physics, Osaka University, Ibaraki, Osaka 567-0047, Japan}
\author{Y.~Fujita}
\affiliation{Department of Physics, Osaka University, Toyonaka, Osaka 567-0043, Japan}
\author{P.~Grabmayr}
\affiliation{Physikalisches Institut, Universit\"{a}t T\"{u}bingen, D-72076 T\"{u}bingen, Germany}
\author{K.~Hatanaka}
\affiliation{Research Center for Nuclear Physics, Osaka University, Ibaraki, Osaka 567-0047, Japan}
\author{D.~Ishikawa}
\affiliation{Research Center for Nuclear Physics, Osaka University, Ibaraki, Osaka 567-0047, Japan}
\author{H.~Matsubara}
\affiliation{Research Center for Nuclear Physics, Osaka University, Ibaraki, Osaka 567-0047, Japan}
\author{Y.~Meada}
\affiliation{Department of Applied Physics, Miyazaki University, Miyazaki 889-2192, Japan}
\author{H.~Okamura}
\affiliation{Research Center for Nuclear Physics, Osaka University, Ibaraki, Osaka 567-0047, Japan}
\author{K.E.~Rehm}
\affiliation{Physics Division, Argonne National Laboratory, Argonne, IL 60439, USA}
\author{Y.~Sakemi}
\affiliation{Cyclotron Radioisotope Center, Tohoku University, Sendai, Miyagi 980-8578, Japan}
\author{Y.~Shimizu}
\affiliation{Center for Nuclear Study, University of Tokyo, Bunkyo, Tokyo 113-0033, Japan}
\author{H.~Shimoda}
\affiliation{Department of Physics, Kyushu University, Higashi, Fukuoka 812-8581, Japan}
\author{K.~Suda}
\affiliation{Research Center for Nuclear Physics, Osaka University, Ibaraki, Osaka 567-0047, Japan}
\author{Y.~Tameshige}
\affiliation{Research Center for Nuclear Physics, Osaka University, Ibaraki, Osaka 567-0047, Japan}
\author{A.~Tamii}
\affiliation{Research Center for Nuclear Physics, Osaka University, Ibaraki, Osaka 567-0047, Japan}
\author{C.~Wrede}
\affiliation{Yale University, New Haven, CT 06520, USA}

% insert abstract here

\date{\today} \begin{abstract}
 The possibility of observing neutrinoless double $\beta$ decay offers the
 opportunity of determining the effective neutrino mass $\it if$ the nuclear
 matrix element were known.  Theoretical calculations are uncertain and the
 occupations of valence orbits by nucleons active in the decay are likely to
 be important.  The occupation of valence proton orbits in the ground states
 of $^{76}$Ge, a candidate for such decay, and $^{76}$Se, the corresponding
 daughter nucleus, were determined by precisely measuring cross sections for
 proton-removing transfer reactions. As in previous work on neutron
 occupations, we find that the Fermi surface for protons is much more
 diffuse than previously thought, and the occupancies of at least three
 orbits change significantly between the two $0^+$ ground states.
 \end{abstract}

 \maketitle

% body of paper here

Major experimental efforts are under way to observe neutrinoless double
$\beta$ decay, an essential step in determining the nature of the neutrino
\cite{Vogel}. If this process were to be observed, it would not only
demonstrate that neutrinos are their own antiparticles, but the rate may
well give the first direct measure of the neutrino mass {\it if} the
corresponding nuclear matrix element can be reliably calculated. For one of
the likely candidates, $^{76}$Ge, theoretical calculations have yielded
answers that vary widely and this uncertainty will be a major obstacle to
deducing the neutrino mass~\cite{Bahcall}.  It is important to explore
experimental data that could help constrain theoretical models. The {\it
difference} in the configuration of nucleons between the initial and final
states (the 0$^+$ ground states of $^{76}$Ge and $^{76}$Se) is a major
ingredient in the matrix element.  We have undertaken several experiments to
better define the knowledge of the two ground-state wave functions and the
difference between them. In previous experiments we have focused on the
difference in neutron configurations, determining the difference in neutron 
valence-orbit occupancies~\cite{gsn} and in correlations~\cite{gept}.  As
was done for neutrons, here we utilize transfer reactions and the 
Macfarlane-French \cite{macfarlane} sum rules to extract the proton
occupancy of states from proton-removing reactions.

These reactions had been studied before \cite{rotb1}, however the two
previous experiments were done at different times and apparently with
slightly different techniques, so that the relative cross sections may have
additional uncertainties.  In addition, the publications do not report cross
sections, just figures showing selected angular distributions and tables of
spectroscopic factors.

\begin{figure}[h]
 \centering
 \includegraphics[scale=0.41]{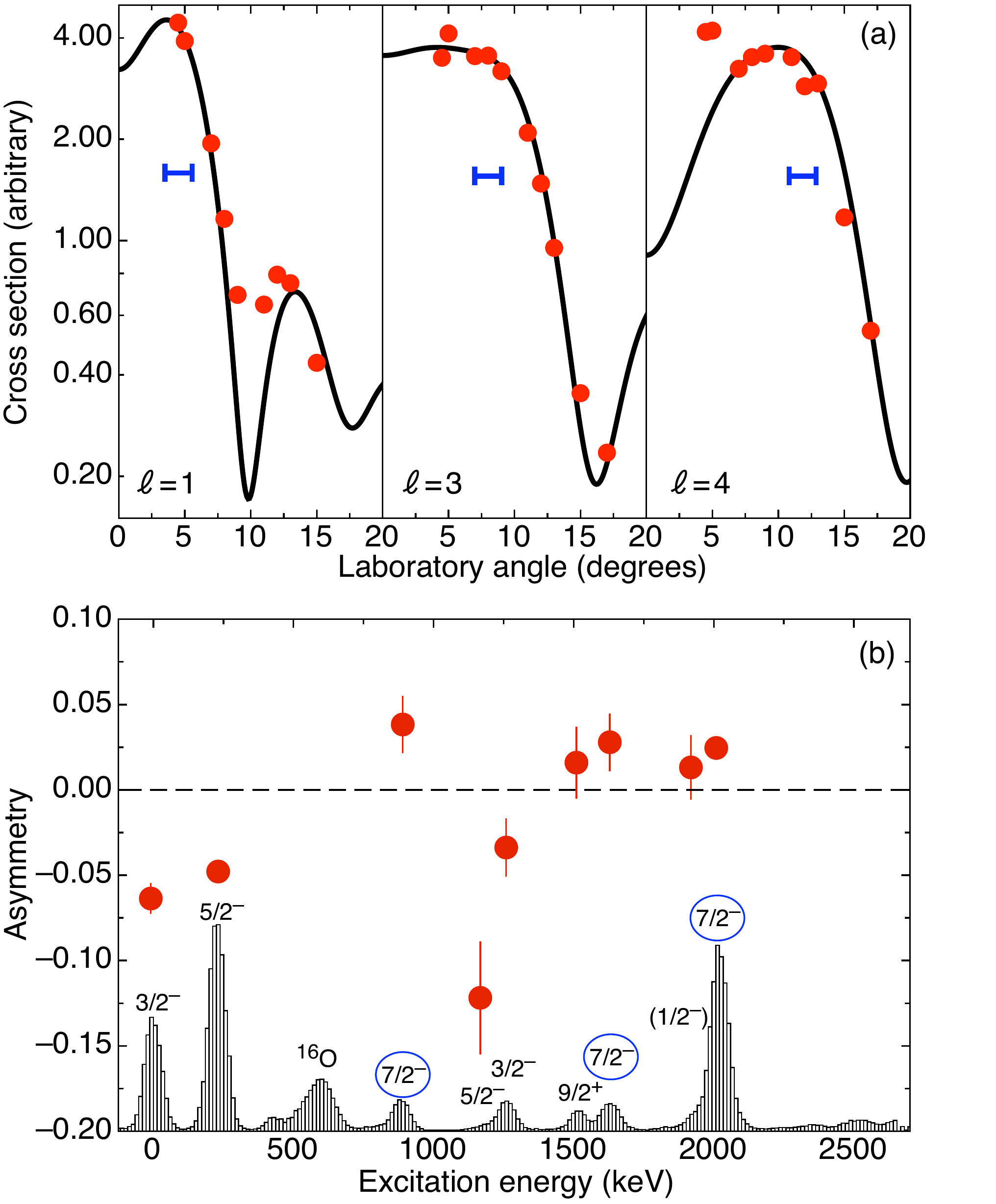}
 \caption{\label{fig1}(color online) (a) Measured angular distributions for $\ell$=1, 3 and 4 on $^{76}$Ge (points) compared to DWBA calculations (lines). The interval shown (blue) indicates the
 angular range covered in the high-statistics measurements. (b) Red points are asymmetries with the vector-polarized deuteron beam,
 between spin up and spin down for $^{76}$Ge, with the data summed over the peaks. A histogram representing the spectrum on the focal plane (in arbitrary units) is inserted at the bottom, with spin assignments shown. The circled spins were determined in this experiment.}
 \end{figure}

There are very few facilities remaining that are capable of carrying out
precision reaction measurements of the type required here, with high-quality
beams, the necessary beam energies, high-resolution spectrometers and
detector systems. We used the RCNP facility at Osaka University to determine
accurate cross sections for the (d,${^3}$He) reaction using the Grand Raiden
(GR) spectrograph~\cite{GR}. Evaporated targets of $^{76}$Ge and $^{76}$Se
were used, as well as targets of $^{74}$Ge and $^{78}$Se as a check on
systematic errors. The experiment was performed at a beam energy of 80 MeV,
the lowest energy feasible, since the focal-plane detector of the GR
requires passage through two foils between the spectrograph vacuum and the
detectors. The angular distributions were found to be consistent with
Distorted Wave Born Approximation (DWBA) calculations, and are shown in
Figure~\ref{fig1}(a). Polarization measurements were also made to distinguish the
spins $j$ of $\ell=3$ transitions, removing uncertainties in attributing
strength to ${f_{5/2}}$ or ${f_{7/2}}$ orbitals.

In order to obtain accurate cross sections, the product of target thickness
and spectrometer aperture was obtained from 10-MeV {$\alpha$} particles at
an angle of 30$^{\circ}$ on the assumption of Rutherford scattering.  The
AVF Cyclotron, usually the injector to the main Ring Cyclotron, was used to
deliver a singly-charged $^4$He beam directly through a bypass beam line, to
the target position.  The spectrometer aperture, 60$\times$40~mrad in the
vertical and horizontal directions, was the same as for all the (d,$^3$He)
measurements with the same current integrator to measure the total beam
charge.

A problem remained regarding the measurements at 4.5$^\circ$.  It is
suspected that the Faraday cup used for the beam measurement at 0$^\circ$
was partially obstructed by components of the spectrometer system. To
appropriately normalize the data, a second spectrograph, the
Large-Acceptance Spectrometer (LAS) placed at a fixed angle of $60^\circ$,
was used at the same time as the GR. The ratio of the scattering yield in
LAS to a beam current integrator was determined when the GR was at larger
angles. This ratio was constant for a given target to the statistical
accuracy of $\sim \pm 3\%$.

%----------------------------FIGURE 1---------------------------------
%---------------------------------------------------------------------

High-statistics measurements of the unpolarized yield were carried out at
the three angles, near the peaks of the angular distributions for $\ell$=1,
3, and 4, at 4.5, 8 and 12$^\circ$, with the aperture subtending $\pm$
1$^\circ$. The choice of the smallest angle was about half a degree larger
than optimal, because of limitations imposed by the counting rates from slit
scattering at the very forward angles.  The operation of the focal-plane
detectors and the method of monitoring efficiencies and dead times have been
described elsewhere~\cite{datahandling}.

The polarization measurements were carried out separately. A
vector-polarized deuteron beam ($A_{y}$=-0.520$\pm$0.010) was used and the
measurement was made with the GR spectrograph at 8$^\circ$, close to optimal
for separating the $j$ values for the $\ell$=3 transitions. Spin assignments
were made to five $\ell$=3 transitions whose $j$ values had not been previously
assigned, and five definite 5/2$^-$ assignments were made, including spins
of the 198-keV level in $^{73}$Ga and the 229-keV level in $^{75}$Ga, both
with dominant spectroscopic factors.  The results for one of the targets,
$^{76}$Ge, are shown in Figure~\ref{fig1}(b).

In order to extract spectroscopic factors, the finite-range DWBA program
PTOLEMY~\cite{ptolemy} was used with several sets of optical-model
parameters from the literature~\cite{opt}. The observed cross sections for
$\ell$=1, 3, and 4, were divided by the DWBA cross sections. A single
overall normalization was established for the (d,$^3$He) data by requiring
that the total occupancy of the valence proton orbits add up to four for Ge
and six for Se.

The DWBA normalizations obtained in this fashion had a mean value of 0.63
with an rms variation of 23\% between the different parameter sets. However,
the summed spectroscopic factors were very nearly the same and varied by
less than 0.1 units.  The values given in Table~\ref{tab1} are the average of results
from analyses with the various potential sets.

We note that `absolute' spectroscopic factors for `good' single-particle
states in doubly-magic nuclei are usually around $\sim$0.6 because of
short-range correlations~\cite{lapikas}. Such correlations are a uniform
property of nuclei, formally they move a given fraction of spectroscopic
strength to very high excitation energy, and this fraction does not change
appreciably between nearby nuclei or configurations. Thus the normalization
obtained by our procedure is meant as a way of estimating the relative
populations of valence orbits---and not as a measurement of absolute
spectroscopic factors.

%----------------------------TABLE 1---------------------------------
\begin{table}
\caption{\label{tab1}Summed spectroscopic strengths.}
\begin{tabular*}{0.474\textwidth}{@{\extracolsep{\fill}}ccccc}
\hline\hline
\multirow{2}{*}{Target}&\multicolumn{4}{c}{Particles}\\
& $\ell$=1 & $\ell$=3 & $\ell$=4 & Sum\\
\hline\\
$^{74}$Ge&1.88&1.52&0.37&3.76\\
{\bf $^{{\bf 76}}$Ge}&{\bf 1.77}&{\bf 2.04}&{\bf 0.23}&{\bf 4.04}\\
{\bf $^{{\bf 76}}$Se}&{\bf 2.08}&{\bf 3.16}&{\bf 0.84}&{\bf 6.08}\\
$^{78}$Se&2.26&1.81&2.20&6.27\\
Uncertainty&$\pm$0.15&$\pm$0.25&$\pm$0.25&$\pm$0.38\\
[1ex]
\hline\hline
\end{tabular*} \end{table}
%---------------------------------------------------------------------

%----------------------------FIGURE 2---------------------------------
\begin{figure}[h]
\centering
\includegraphics[scale=0.43]{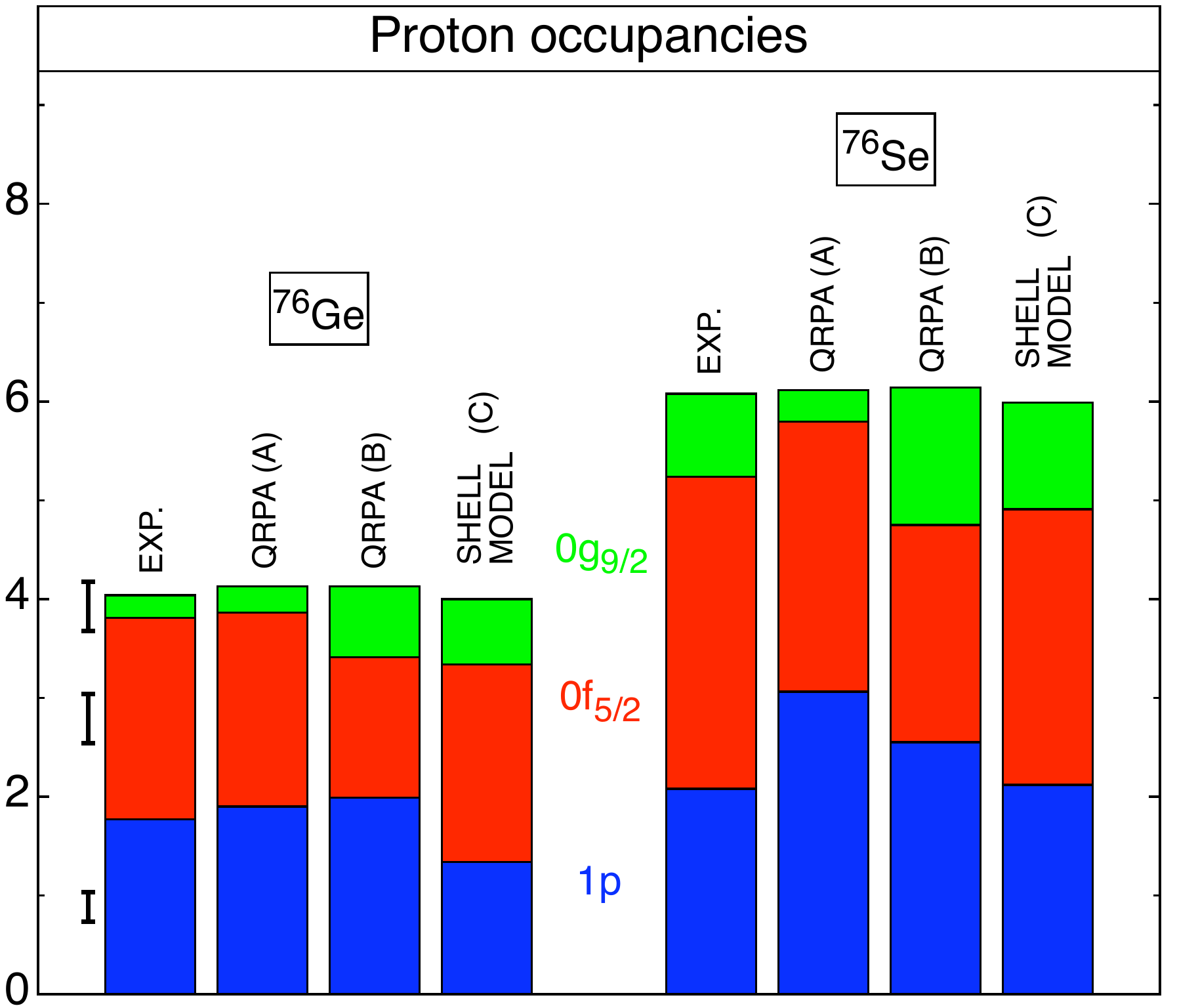}
\caption{\label{fig2}(color online) Experimentally determined proton occupancies from Table~\ref{tab1} for the three valence  orbits in $^{76}$Ge and $^{76}$Se. These are compared to (A) QRPA calculations of Ref.~\cite{rodin}, (B) of
 Ref.~\cite{suhonen} and (C) the shell-model calculations of Ref.~\cite{caurier}. The experimental uncertainties are indicated by
 error bars on the left.}
 \end{figure}
%---------------------------------------------------------------------

The summed spectroscopic factors of Table~\ref{tab1} are the proton occupancies and
are shown graphically in Figure~\ref{fig2}, for the $A=76$ nuclei, along with the
values of these occupancies from various theoretical calculations.  The
first QRPA calculations \cite{rodin} were the only ones available at the
time the neutron occupancies were published \cite{gsn} and were also shown
there. The more recent QRPA calculation \cite{suhonen} and a shell-model
calculation \cite{caurier} were published subsequently.

The uncertainties in the experimental values are difficult to estimate.
Statistical errors in the summed strengths are less than 1\% and relative
systematic errors between targets are believed to be less than 3\%. The
biggest uncertainties stem from possible missing states and, to a lesser
extent, from the validity of the DWBA method in extracting spectroscopic
factors. We estimate that the occupancy is determined to about 0.15 nucleons
for the $1p$, and to 0.25 for the $0f_{5/2}$ and $0g_{9/2}$ orbits. These
estimates of uncertainties are of necessity rather crude -- we have some
confidence in them because a single normalization gives the appropriate
proton number for the four targets studied, with an rms deviation of
$\pm$0.18 nucleons.

%----------------------------FIGURE 3---------------------------------
\begin{figure}[h]
\centering
\includegraphics[scale=0.53]{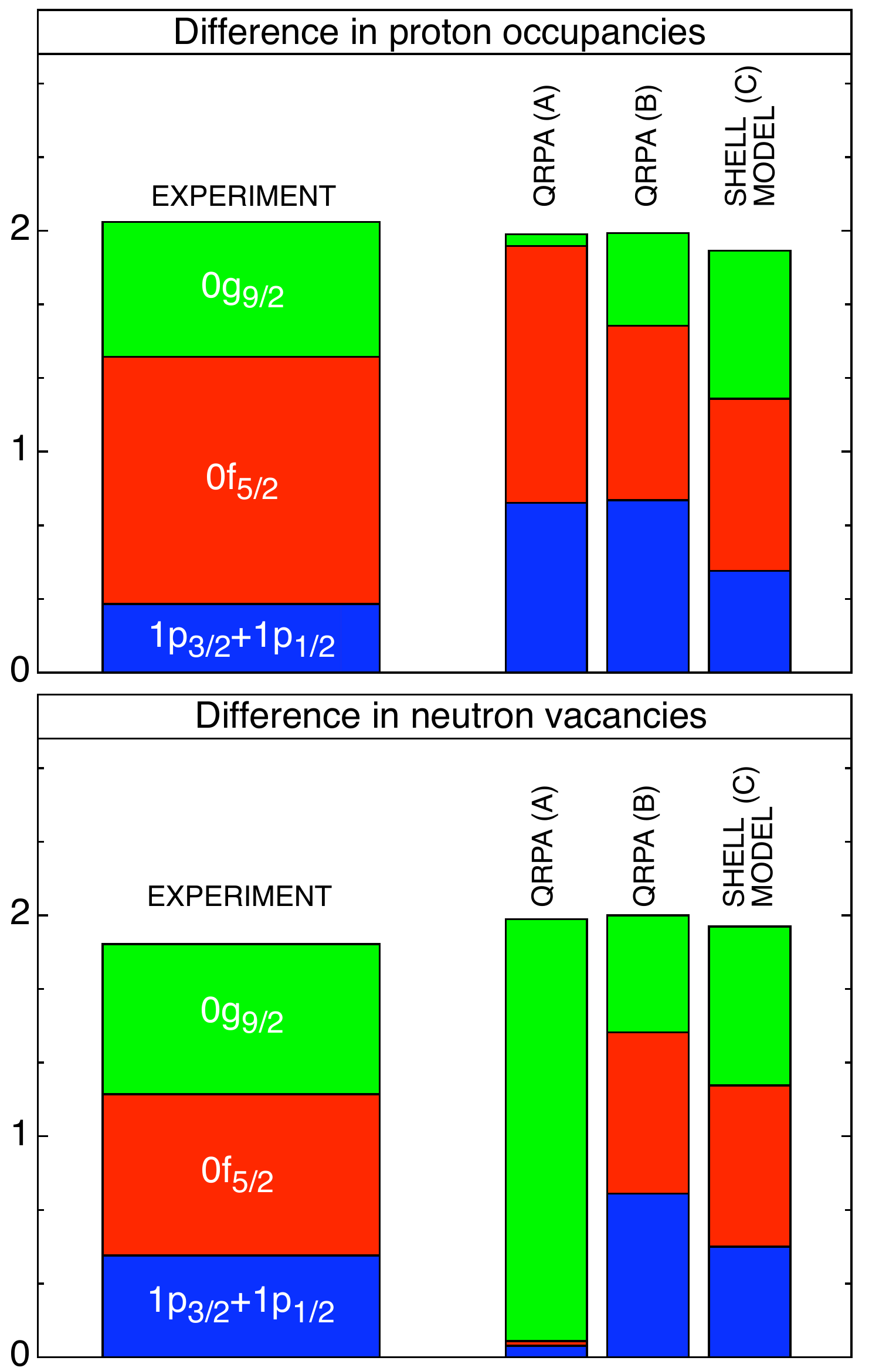}
\caption{\label{fig3}(color online) The {\it difference} between proton and neutron~\cite{gsn} occupancies of the ground states of $^{76}$Ge and $^{76}$Se deduced from the present measurement and compared to theoretical calculations with the notation as in Fig.~\ref{fig2}.}
\end{figure}
%---------------------------------------------------------------------

We also made measurements of the ($^3$He,d) reaction on the same targets and
at the same scattering angles, at a corresponding $^3$He beam energy of 72
MeV---determined by the average difference in $Q$-values between the
proton-adding and proton-removing reactions on these targets.  Even though
Ge has only 32 protons, and thus the major shell between $Z=28$ and $50$ is
barely started, we intended to attempt to apply the sum rule to the
proton-adding reaction as well as to the proton-removing one.

However, when we apply the same normalization to the ($^3$He,d) data, unlike
the neutron case, the summed strengths for proton vacancies fall short of
the expected values, indicating considerable missed strength.  For $\ell$=1
an average of 16\% of the total strength (12, 14, 21 and 18 \% respectively
for $^{74}$Ge, $^{76}$Ge, $^{76}$Se, and $^{78}$Se), for $\ell$=3 24\% (25,
30, 4, and 35), and for $\ell$=4 48\% (56, 41, 57, 37) are missing.  A small
correction, for the unobserved T$_>$ strength was made in the sums --- these
strengths are known from the measurements of the neutron vacancies.

The shortfall may be a consequence of the fact that these nuclei are near
the beginning of a major shell for protons, the centroids of the expected
single-particle strength are at higher excitation energy and the strength
becomes appreciably fragmented in a high level-density region such that the
many small individual components cannot be resolved. Indeed, the discrepancy
is worst for the 9/2$^+$ orbit, which is at the highest excitation energy.
Previous work on the ($^3$He,d) reaction at lower bombarding
energies~\cite{he3d}, and some of it at higher energy resolution, has not
identified significant components with the requisite $\ell$ values. We have
also attempted to identify this missing strength by carrying out
measurements at the Yale ESTU tandem of the ($\alpha$,t) reaction which is
particularly selective for $\ell$=4 transitions, where the problem is the
most serious. No candidates for missing $\ell$=4 strength were found with a
strength greater than $\sim$ 4\%. This is unlike the neutron-transfer case,
where excellent consistency was obtained between the neutron-adding and
removing reactions with $\ell$=4 using the ($\alpha$,$^{3}$He) reaction and
its inverse.  We therefore conclude that we cannot utilize the data from
proton-adding reactions in the present case and must rely on the consistency
of the proton-removing reactions.

Indeed, the strength in the proton-removing reactions so close to the $Z=28$
closed shell, is restricted to relatively few states at low excitation
energies and, as is shown in Table~\ref{tab1}, gives self-consistent results among
the four nuclei studied.  Even though the summed strength was used for the
normalization, the four determinations are independent and the rms deviation
from the expected value is 0.19 protons -- well within the estimated
uncertainty.  We believe that, together with the neutron occupation study
\cite{gsn}, the internal consistency of these results perhaps constitutes
the most quantitative test of the validity of the sum rules in one-nucleon
transfer reactions.

Utilizing the information in Table~\ref{tab1}, we can state the differences in proton
occupations between $^{76}$Se and $^{76}$Ge as 0.31$\pm$0.15 in $\ell$=1 (it
appears that this difference is mostly in the ${0p_{1/2}}$ occupation),
1.12$\pm$0.35 in ${0f_{5/2}}$, and 0.61$\pm$0.35 in ${0g_{9/2}}$. The
uncertainties were estimated taking possible correlations into account. This
comparison is shown in Figure~\ref{fig3}, displaying both the neutron differences
from \cite{gsn} and the proton differences from the present work, along with
the same set of calculations as in Figure~\ref{fig2}.  As is evident from both
Figures~\ref{fig2} and~\ref{fig3}, the $0g_{9/2}$ proton orbit is considerably more involved
in the changes in the Fermi surface than the original QRPA \cite{rodin}
calculations suggested.  The more recent calculations are consistent with
the proton as well as the neutron data on occupancies.  Some weak $\ell$=2
transitions are also seen in the proton removal from the two Se isotopes.
The spectroscopic factors indicate $1d_{5/2}$ strength of no more than 0.1
protons. Low-lying 7/2$^-$ states seen in the ($^3$He,d) reaction indicate
similarly low strengths for these transitions.  Though this evidence is not
conclusive -- we see no appreciable admixture from beyond $Z=50$ or below
$Z=28$.

We have thus characterized the microscopic changes in the valence
occupations for neutrinoless double beta decay.  Measurements of neutron
pair-transfer on these nuclei suggest that correlations between neutrons are
very similar in these two ground states~\cite{gept}.  We hope to determine
proton correlations in a future experiment.  Although the most recent
theoretical calculations by~\cite{suhonen},~\cite{caurier} and~\cite{vogel} appear
to be closer to each other in their predictions for the $^{76}$Ge matrix
element, firm anchor points to the measured properties of these states are
likely to be essential. At present, the theoretical calculations do not yet
seem to be be able to specify how the experimental observables that
characterize the wave functions of these two ground states influence the
matrix element for neutrinoless double $\beta$ decay.

We are indebted to John Greene for preparing targets for these experiments.
This measurement (E292) was performed at RCNP, Osaka University. The authors
wish to thank the RCNP operating staff, and the participants from outside
RCNP wish to thank the local staff and administration for their hospitality
and assistance.  The work was supported by the U.S. Department of Energy,
Office of Nuclear Physics, under contracts DE-FG02-91ER-40609 and
DE-AC02-06CH11357, the UK Science and Technology Facilities Council, and the
German BMBF.

Our data are available on-line in the XUNDL database \cite{xundl}.

\end{document}